\documentclass[aip,amsmath,amssymb,reprint]{revtex4-1}

\usepackage{graphicx}   % Include figure files
\usepackage{dcolumn}    % Align table columns on decimal point
\usepackage{bm}         % Bold math
\usepackage{color}      % Color support
\usepackage{hyperref}   % Hyperlinks
\usepackage{url}
\usepackage{hyperref}
\begin{document}

\preprint{AIP/POF}  % Optional: remove or modify

%\preprint{}

\title{Investigating frictional instability due to pressurization in granular media: insights from coupled computational
fluid dynamics–discrete element method} %Title of paper



\author{Bimal Chhushyabaga}
\email[]{bchhushyabaga@uh.edu}
%\thanks{}
\affiliation{Department of Civil and Environmental Engineering, University of Houston}

\author{Behrooz Ferdowsi}
\email[]{behrooz@Central.UH.EDU}
%\thanks{}
\affiliation{Department of Civil and Environmental Engineering, University of Houston}


\date{\today}

\begin{abstract}
Fluid pressurization can reactivate subcritically stressed granular layers in faults, slopes, and injection-perturbed reservoirs, yet the grain-scale feedbacks linking pressure diffusion, drainage, and contact-network degradation to frictional instability remain difficult to resolve. Here, we investigate pore-pressure-induced reactivation of confined, fluid-saturated granular shear layers under prescribed shear stress using three-dimensional coupled computational fluid dynamics–discrete element method (CFD–DEM) simulations. Strain-controlled tests first define the Mohr–Coulomb strength envelope; stress-controlled simulations then impose subcritical shear stresses while basal pore pressure is increased under drained and undrained boundary conditions. The simulations show instability is not governed by pore pressure alone, but by its coupled evolution with effective stress, drainage, dilation or compaction, hydraulic connectivity, and granular fabric. Undrained boundaries retain excess pore pressure, whereas drained boundaries maintain vertical pressure gradients and suppress excess pressure. Internal fields reveal alternating dilation and compaction bands and reorganization of a porosity-derived permeability proxy, indicating that hydraulic pathways evolve during deformation. Micromechanical diagnostics identify localized particle rotation, force-chain reorganization, porosity redistribution, and coordination-number variations controlled mainly by imposed shear-stress level rather than drainage. Second-order fabric metrics show that post-failure weakening is accompanied by loss of directional force-chain organization, especially in the lower-shear case. Friction–velocity and friction–porosity trajectories indicate a transition from dilatancy-dominated strengthening to pore-pressure-driven weakening. Viscous-number scaling partially organizes the low-$I_v$ creeping response,  $10^{-8}\lesssim I_v\lesssim10^{-5}$, but does not collapse onto a unique local rheology. These results clarify how drainage-controlled hydromechanical feedbacks and fabric degradation convert pore-pressure forcing into frictional instability.
\end{abstract}


\maketitle %\maketitle must follow title, authors, abstract and \pacs


\section{Introduction}

Fluid pressurization is a primary trigger of deformation and instability in saturated geomaterials, including fault gouge, rainfall-affected slopes, creeping landslides, injection-perturbed reservoirs, and subaqueous granular deposits \citep{Iverson2000,Iverson2005,Cappa2014,Guglielmi2015,Agliardi2020,Ellsworth2013}. These systems often contain localized or incipient shear zones that are subcritically stressed, so a moderate pore-pressure increase can reduce effective normal stress and bring the material toward failure while the applied shear stress remains nearly constant \citep{HubbertRubey1959,Terzaghi1943,Nguyen2021}. In a Mohr--Coulomb framework, this effect is expressed as
\begin{equation}
\tau_c=c+\mu(\sigma_n-p),
\end{equation}
where $\tau_c$ is the critical shear stress, $c$ is cohesion, $\mu$ is the friction coefficient, $\sigma_n$ is total normal stress, and $p$ is pore pressure. This relation provides a useful macroscopic failure criterion, but it does not explain how deformation evolves before failure, why pressurized systems may creep stably or accelerate dynamically, or how grain-scale processes control the transition to instability \citep{Scuderi2017,Cappa2019,Nguyen2021,Sarma2025}.

From a fluid-mechanical perspective, the transition from stable deformation to rapid weakening is controlled by imposed pressure forcing, pore-pressure diffusion, and deformation-induced pore-volume change \citep{SegallRice1995,Segall2010,Sarma2025}. For a diffusion-controlled response, the relevant drainage time can be estimated as $t_d \sim L_d^2/D_h$, where $L_d$ is the drainage length and $D_h$ is an effective hydraulic diffusivity \citep{RiceCleary1976,DetournayCheng1993}. Pressurization over time scales shorter than $t_d$ favors undrained pressure retention, effective-confinement loss, and abrupt weakening, whereas efficient drainage maintains pressure gradients and limits excess-pressure buildup \citep{RiceCleary1976,DetournayCheng1993,Scuderi2017,Cappa2019}. Because shear-induced dilation, compaction, and porosity reorganization modify both pore pressure and hydraulic connectivity, the drainage response can evolve during failure nucleation \citep{SegallRice1995,RudnickiChen1988,Nguyen2021,Sarma2025}. This coupling among pressurization, diffusion, drainage, and granular deformation motivates the drained--undrained comparison used in this study.

The limitation of a purely macroscopic criterion is especially important for granular shear zones, where strength and stability emerge from evolving particle-scale structure. In sheared gouge and granular layers, deformation may transition from distributed shear to localized slip zones, modifying frictional strength, porosity, permeability, and pore-pressure diffusion pathways \citep{Sulem1995,Marone1998,RathbunMarone2010,RathbunMarone2013}. At the grain scale, shear resistance is controlled by particle rearrangement, rolling and sliding contacts, coordination number, fabric anisotropy, and the organization of strong force chains \citep{Jaeger1996,Rothenburg1989,Radjai1998,Roux2002,Tordesillas2007}. DEM studies further show that localization and stick--slip behavior depend on the evolving contact network, grain friction, particle shape, wall roughness, and loading condition \citep{CundallStrack1979,MorganBoettcher1999,AharonovSparks2004,MairHazzard2007,RathbunRenardAbe2013}. Thus, the stability of a pressurized granular layer depends not only on effective stress but also on the integrity of the load-bearing contact skeleton.

Pore fluids add a second feedback by coupling volumetric deformation to pore-pressure evolution. Dense granular materials often dilate during shear, and dilation can reduce pore pressure, increase effective stress, and stabilize slip through dilatant hardening \citep{Reynolds1885,Frank1965,Scholz1973,SegallRice1995,RudnickiChen1988}. Conversely, compaction, rapid undrained loading, thermal pressurization, or external injection can elevate pore pressure and promote weakening when pressure generation exceeds hydraulic diffusion or drainage \citep{Lachenbruch1980,Rice2006,Proctor2020,Brantut2020}. Failure therefore reflects a competition among pore-pressure increase, dilation-induced strengthening, compaction-induced pressurization, permeability evolution, and stress redistribution \citep{Segall2010,Scuderi2017,Cappa2019,Sarma2025}. This competition can produce delayed failure, arrested preslip, frictional hysteresis, rate-dependent strengthening, and transitions from slow creep to rapid slip in saturated granular media \citep{pailha2008initiation,PailhaPouliquen2009,Nguyen2021,Sarma2025}.

Laboratory, field, and continuum studies have established that pore-pressure perturbations can induce both aseismic and dynamic slip depending on drainage, loading rate, stiffness, material state, and stress proximity to failure \citep{Cornet1997,Guglielmi2015,ScuderiCollettini2016,Scuderi2017,Cappa2019}. Continuum and reduced-order hydromechanical models represent these processes through poroelastic diffusion, Darcy-type flow, porosity-dependent permeability, frictional constitutive laws, and volumetric-strain or dilatancy source terms \citep{RiceCleary1976,DetournayCheng1993,SegallRice1995,Segall2010,Segall2015,Cappa2018,Cappa2019}. These models are essential for field-scale interpretation, but their closures become increasingly uncertain near failure, when pore pressure, porosity, strain rate, contact forces, and granular fabric evolve heterogeneously.

Capturing these near-failure feedbacks requires a grain-resolving, two-way coupled fluid--particle framework in which the fluid and solid phases exchange momentum while retaining the evolving granular fabric. A CFD--DEM formulation is well suited for this problem because the pore fluid acts on particles through pressure-gradient forces and Syamlal--O'Brien hindered drag, while particle motion changes the local fluid volume fraction and modifies the pathways available for pressure redistribution \citep{Syamlal1993,MFIXDEM2012,Musser2021}. Coupled CFD--DEM simulations therefore provide a complementary route for linking pore-pressure diffusion, drainage condition, fluid--solid momentum exchange, contact-network degradation, and macroscopic slip behavior \citep{Goren2011,Dorostkar2017,YangJuanes2018,Nguyen2021,Sarma2025}.

Grain-resolving numerical methods provide a complementary route for linking macroscopic slip behavior to micromechanical mechanisms. DEM has been widely used to study granular gouge friction, stick--slip, strain localization, force-chain dynamics, and frictional healing in dry granular systems \citep{CundallStrack1979,AharonovSparks2004,Ferdowsi2013,FerdowsiRubin2020,Parez2021}. Coupled CFD--DEM methods extend this capability to saturated media by linking particle motion to pore-pressure evolution and fluid--solid momentum exchange \citep{Goren2011,Dorostkar2017,YangJuanes2018,Nguyen2021,Sarma2025}. Recent coupled simulations show that fluid pressurization can drive slow creep before failure and dynamic slip afterward, with the transition controlled by localization, particle rotation, and force-chain collapse \citep{Nguyen2021}; pressure cycling in saturated gouge can also produce delayed rupture, dilatant preslip, hysteresis, and shear-rate strengthening \citep{Sarma2025}. However, the combined roles of drainage condition, pressurization rate, initial stress level, pore-pressure redistribution, hydraulic-connectivity evolution, and contact-network degradation remain incompletely resolved in three-dimensional CFD--DEM simulations. 

The present study is distinct from our previous CFD–DEM investigation of fluid-saturated granular collapse and runout \citep{Chhushyabaga2025}. That work focused on dense and loose granular assemblies under subaerial and subaqueous conditions, emphasizing how initial packing state, dilation or compaction, pore-pressure evolution, and inertial or viscous rheological organization control collapse dynamics and runout. Here, we address a different physical problem: pore-pressure-induced reactivation of an initially confined granular shear layer under prescribed shear stress. The novelty of the present work lies in resolving how a subcritically stressed, fluid-saturated granular layer transitions from stable deformation to instability as basal pore pressure is increased, and how this transition is regulated by drainage condition, pressurization rate, effective-stress reduction, dilation or compaction, hydraulic-connectivity evolution, and degradation of the force-bearing contact network.

The loading protocol is central to resolving these mechanisms. Strain- or velocity-controlled shear is useful for measuring frictional strength, rate dependence, steady-state behavior, and Mohr--Coulomb parameters \citep{Marone1998,RathbunMarone2010}. However, prescribing displacement can suppress the natural acceleration that defines instability. Stress-controlled loading instead allows slip velocity, dilation or compaction, localization, pore-pressure evolution, and contact-network degradation to emerge from internal feedbacks \citep{Scuderi2017,Cappa2019,Nguyen2021,Sarma2025}. This distinction is directly relevant to pressurized slopes, reservoirs, and faults, where failure occurs when the evolving effective stress and granular strength can no longer support the imposed shear load.

Here, we use a two-way coupled, volume-averaged CFD--DEM framework to investigate pressurization-driven failure in a fluid-saturated granular layer. Strain-controlled simulations under servo-controlled normal stress first establish the reference frictional strength and Mohr--Coulomb failure envelope. Stress-controlled simulations then impose shear stresses below the reference strength while pore pressure is progressively increased from the base under drained and undrained hydraulic boundaries. We analyze apparent friction, slip velocity, excess pore pressure, dilatancy, porosity, and a porosity-derived effective-permeability together with particle rotation, force-chain reorganization, coordination-number evolution, and porosity redistribution. Coordination number is used as a bulk measure of contact-network connectivity, whereas force-chain maps and particle-rotation fields provide information on the strength, organization, and localization of the load-bearing fabric. This distinction is important because drainage may strongly alter pore-pressure redistribution and effective-stress evolution even when the bulk coordination number remains similar between drained and undrained cases. 

To connect the spatial force-chain observations to quantitative fabric evolution, we also evaluate second-order contact and force-chain anisotropy metrics following the stress--force--fabric interpretation of granular shear resistance \citep{Rothenburg1989}. These metrics distinguish the anisotropy of the full contact network, the force-weighted load-bearing fabric, and the strong-force backbone, allowing the directional degradation of the granular skeleton to be compared across stress states. To place the simulated response in the context of dense-suspension and immersed-granular rheology, we also compare the frictional response with viscous-number scaling. In pressure-imposed suspensions, the friction coefficient can often be organized using the viscous number $I_v=\eta_f\dot{\gamma}/P'$, where $\eta_f$ is the pore-fluid viscosity, $\dot{\gamma}$ is the representative shear strain rate, and $P'$ is the effective particle pressure \citep{Boyer2011}. This dimensionless number compares viscous shear forcing with effective particle pressure. Sediment-bed experiments further show that local $\mu(I_v)$ rheology can describe mobile regions but breaks down at very low $I_v$, where creep, intermittent rearrangement, and nonlocal contact-network effects become important \citep{Houssais2016}. We therefore use viscous-number scaling to evaluate the rheological organization of the transient pressurization-driven response. The objective is to determine how pore-pressure buildup, volumetric deformation, hydraulic-connectivity changes, and granular-fabric degradation interact during instability, and how drainage regulates the conversion of pore-pressure forcing into effective-stress reduction, contact-network failure, and slip acceleration. The results provide a grain-resolving insights for interpreting pore-pressure-induced weakening in fault gouge, rainfall-triggered slopes, and injection-perturbed granular systems.

\begin{figure*}
  \centering
  \includegraphics[width=1.0\textwidth]{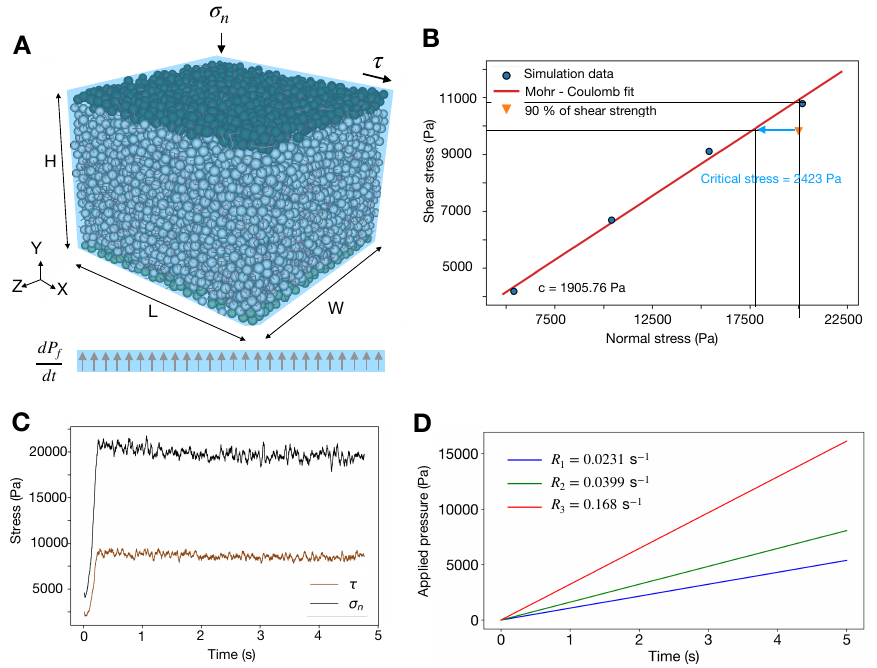}
  \caption{Numerical model geometry, loading framework used in the CFD--DEM simulations. (a) Simple-shear configuration of a fluid-saturated granular layer confined between a fixed bottom plate and a movable top plate. The model dimensions are $L=50d_p\approx0.05$~m, $W=50d_p\approx0.05$~m, and $H=38d_p\approx0.038$~m. A constant normal stress $\sigma_n=20$~kPa is applied at the top boundary, while shear is imposed through a prescribed shear-stress $\tau$. The three imposed shear-stress levels used in the stress-controlled pressurization stage are defined relative to the reference steady-state shear strength $\tau_{ss}$, such that $\tau_1=0.9\tau_{ss}$, $\tau_2=0.68\tau_{ss}$, and $\tau_3=0.46\tau_{ss}$. Fluid pressurization is applied from the base at a rate $\mathrm{d}P_f/\mathrm{d}t$. (b) Mohr--Coulomb strength calibration obtained from strain-controlled simulations at different normal stresses; the fitted failure envelope yields cohesion $c=1905.76$~Pa and a critical shear stress of 2423~Pa to initiate failure.  (c) Representative time evolution of the applied normal and shear stresses in a stress-controlled simulation, showing regulation of $\sigma_n$ and $\tau$ during shearing. (d) Imposed base pressurization histories for the three normalized pressurization rates, with $\mathrm{d}P_f/\mathrm{d}t=462$, 798, and 3360~Pa/s, corresponding to $R_1=0.0231$~s$^{-1}$, $R_2=0.0399$~s$^{-1}$, and $R_3=0.168$~s$^{-1}$, respectively.}
  \label{fig:Model_Setup}
\end{figure*}

\begin{figure*}[t]
  \centering
  \includegraphics[width=0.9\textwidth]{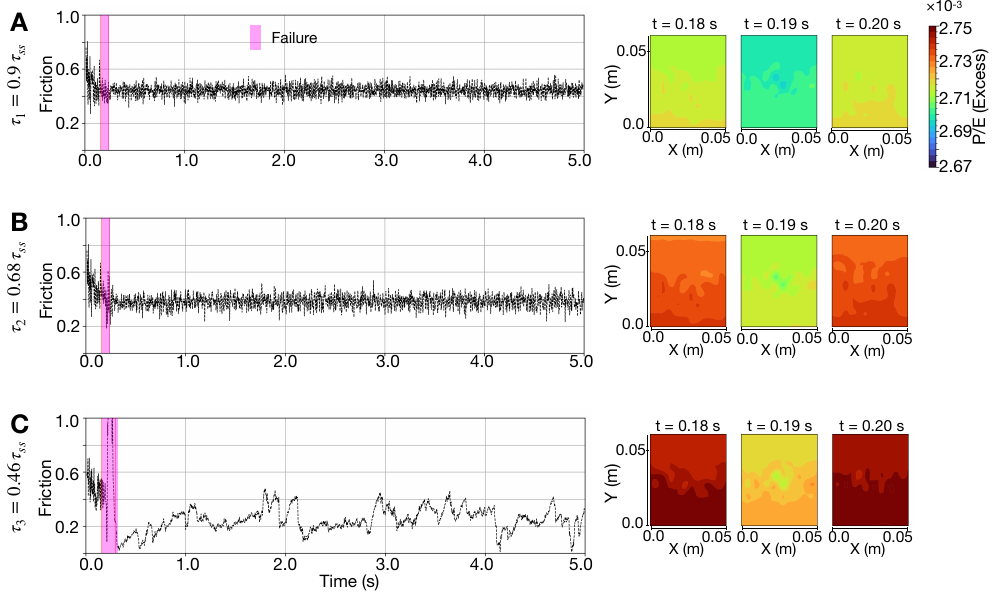}
  \caption{Undrained pressurization-driven response of the granular layer at the highest pressurization rate, $R_3$, for three imposed shear-stress levels relative to the reference shear strength, $\tau_1=0.9\tau_{ss}$, $\tau_2=0.68\tau_{ss}$, and $\tau_3=0.46\tau_{ss}$. Panels A--C show the temporal evolution of the apparent friction coefficient, $\mu=\tau/(\sigma_n-p_f)$, over the full simulation time, together with the identified failure interval highlighted in magenta. It marks the objectively identified failure window, defined from the local peak in $\mu_{\rm app}$ to the subsequent local minimum when the friction drop satisfies $\Delta\mu=\mu_{\rm app}^{\rm peak}-\mu_{\rm app}^{\rm min}\ge0.01$, following the friction-drop criterion used by Dorostkar et al.~\citep{Dorostkar2017}. The corresponding maps on the right show the spatial distribution of normalized excess pore pressure, $P/E$, at three representative stages spanning failure onset, peak pressurization, and post-failure pressure release ($t=0.18$, 0.19, and 0.20~s). Where,  $E=100~MPa$ is particle Young's modulus. The comparison demonstrates that the highest imposed shear-stress case remains closest to failure and requires only a modest pore-pressure increase to weaken, whereas the lowest imposed shear-stress case exhibits the largest friction drop and the strongest transient pore-pressure buildup before instability. Across all three cases, failure is marked by abrupt frictional weakening with localized excess pore-pressure.}
  \label{fig:Fric_pres_un}
\end{figure*}

\begin{figure*}[t]
  \centering
  \includegraphics[width=0.9\textwidth]{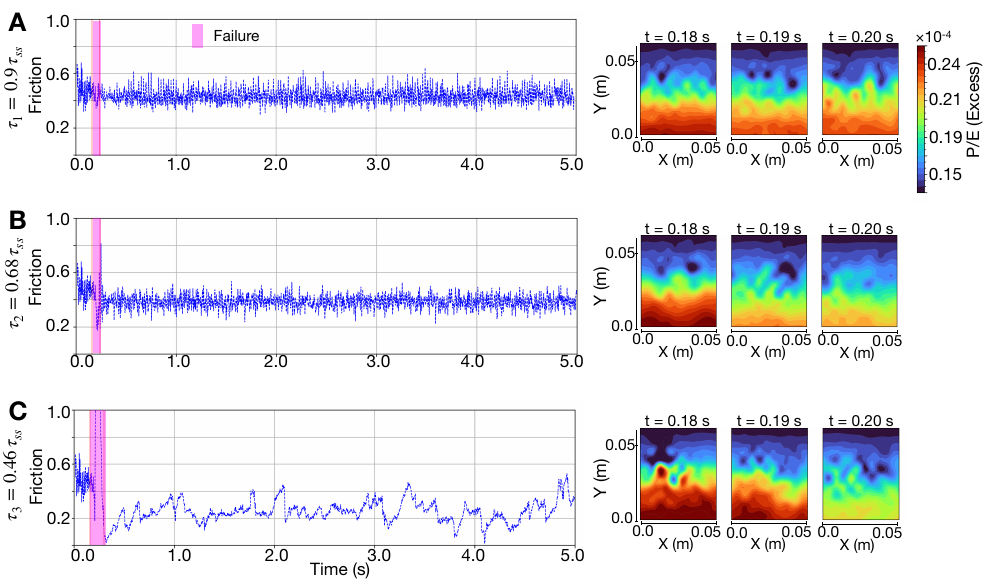}
  \caption{Drained pressurization-driven response of the granular layer at the highest pressurization rate, $R_3$, for three imposed shear-stress levels relative to the reference shear strength, $\tau_1=0.9\tau_{ss}$, $\tau_2=0.68\tau_{ss}$, and $\tau_3=0.46\tau_{ss}$. Panels A--C show the temporal evolution of the apparent friction coefficient, $\mu=\tau/(\sigma_n-p_f)$, over the full simulation time, together with the identified failure interval highlighted in magenta. It denotes the failure interval selected using the same friction-drop criterion, $\Delta\mu\ge0.01$, with onset at the local maximum in $\mu_{\rm app}$ before sustained weakening and termination at the subsequent post-drop minimum~\citep{Dorostkar2017}. The corresponding maps on the right show the spatial distribution of normalized excess pore pressure, $P/E$, at three representative stages spanning the onset of weakening and subsequent evolution ($t=0.18$, 0.19, and 0.20~s). Where,  $E=100~MPa$ is particle Young's modulus. Relative to the undrained case, drainage strongly suppresses excess pore-pressure buildup, as reflected by the substantially smaller pressure scale and the persistent vertical pressure gradient from the pressurized base toward the drained top boundary. }
  \label{fig:Fric_pres_d}
\end{figure*}

\begin{figure*}[t]
  \centering
  \includegraphics[width=1.0\textwidth]{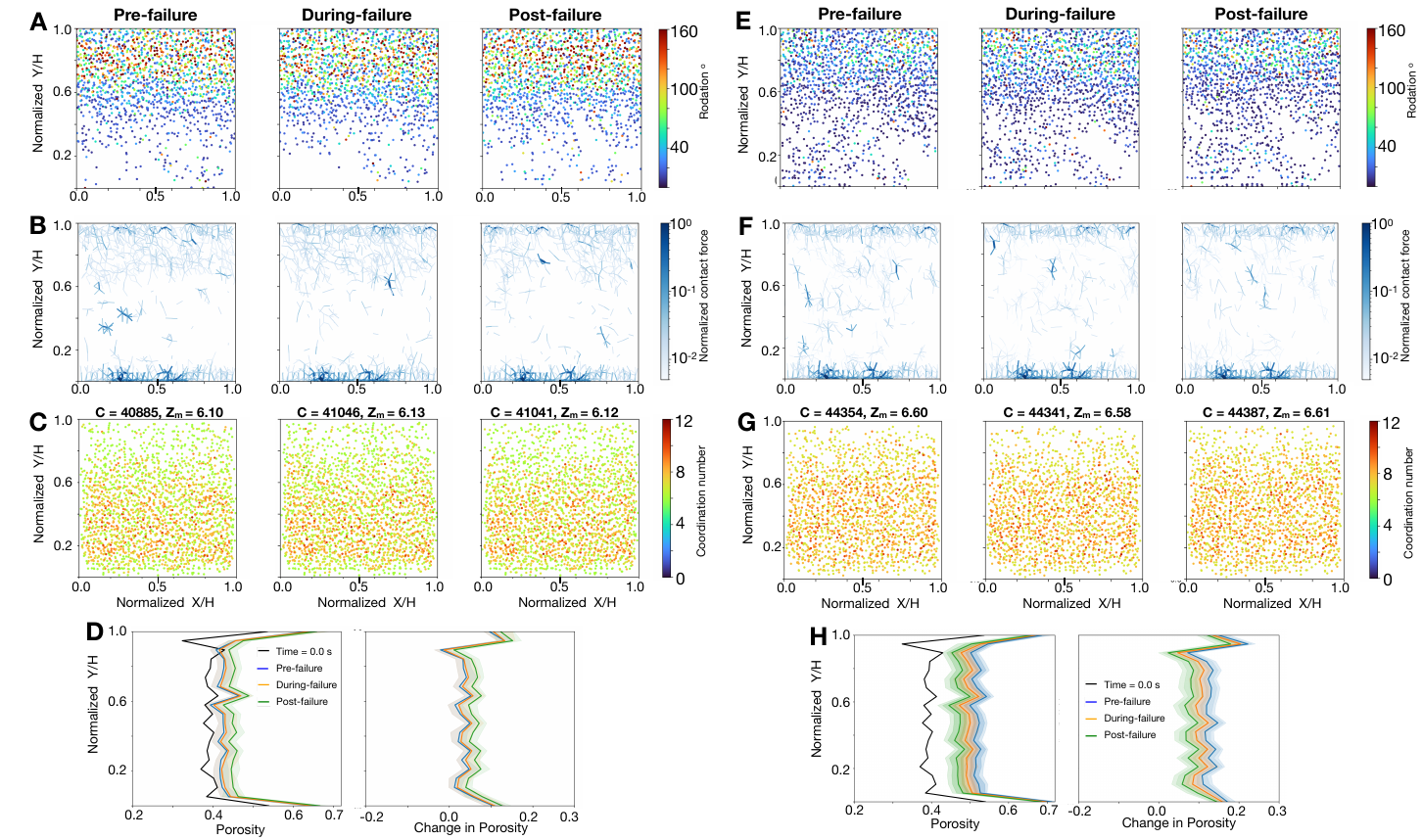}
\caption{Micromechanical evolution of the undrained granular layer before, during, and after failure for two representative pressurization-driven cases: R3--T1 (panels A--D; left three columns) and R3--T3 (panels E--H; right three columns). Within each case, columns correspond to pre-failure, during-failure, and post-failure states. From top to bottom, the rows show: particle rotation magnitude (A,E), force-chain structure (B,F), coordination number (C,G), and normalized porosity profiles with corresponding change in porosity relative to the initial state (D,H). In the first row, particles are colored by rotation magnitude, highlighting increasingly localized rotational activity during failure. In the second row, line color and thickness scale with normalized contact-force magnitude on a logarithmic scale, showing degradation and reorganization of the strong-force backbone across the failure sequence. In the third row, particles are colored by coordination number; shows total number of contacts $C$ and coordination number $Z_m$. In the fourth row, porosity and change-in-porosity profiles are plotted as functions of normalized height, comparing the initial state at $t=0.0$ s with the pre-failure, during-failure, and post-failure states. All spatial maps are plotted in normalized coordinates $X/H$ and $Y/H$. The figure shows that failure is accompanied by concentrated particle rotation, collapse and redistribution of the contact-force, changes in contact connectivity, and porosity redistribution, with systematically different micromechanical responses between the higher-shear case (R3--T1) and the lower-shear case (R3--T3).}
  \label{fig:Micro_Mechanics}
\end{figure*}

\begin{figure*}[t]
  \centering
  \includegraphics[width=1.0\textwidth]{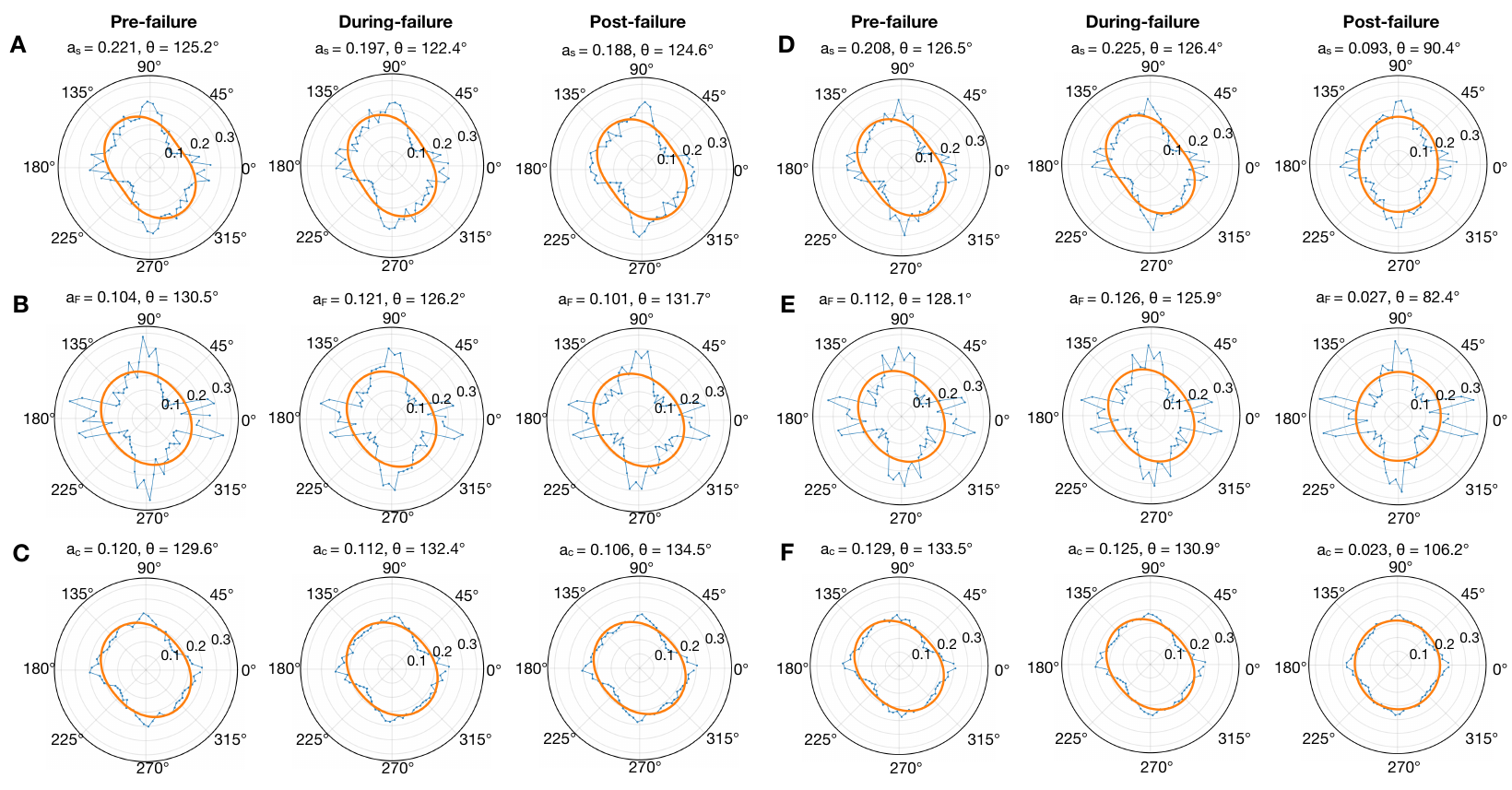}
  \caption{Directional evolution of force-chain and contact-fabric anisotropy for the representative undrained R3--T1 and R3--T3 cases. Panels A--C show R3--T1, the case closest to the reference shear strength, $\tau_1=0.90\tau_{ss}$, and panels D--F show R3--T3, the lowest-shear case, $\tau_3=0.46\tau_{ss}$, at the same pressurization rate $R_3$. Within each case, columns correspond to pre-failure, during-failure, and post-failure states. From top to bottom, rows show the strong-contact anisotropy $a_s$ (A,D), force-weighted fabric anisotropy $a_F$ (B,E), and contact-fabric anisotropy $a_c$ (C,F). Blue curves denote normalized directional distributions, and orange curves denote second-Fourier anisotropy fits. The distributions are plotted in a full $0$--$2\pi$ representation for visualization, while the anisotropy coefficients are computed from the equivalent bidirectional $0$--$\pi$ contact-orientation distributions. R3--T3 exhibits a strong post-failure reduction in $a_s$, $a_F$, and $a_c$, indicating substantial degradation of the directional contact and force-bearing fabric, whereas R3--T1 retains a more persistent anisotropic load-bearing structure.}

  \label{fig:Force_Fabric_Anisotropy}
\end{figure*}

\begin{figure}[t]
  \centering
  \includegraphics[width=\columnwidth]{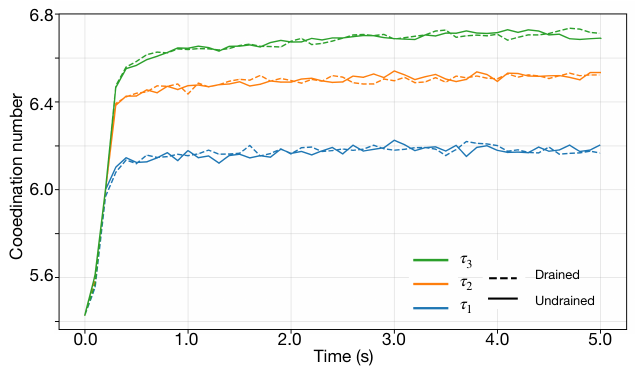}
  \caption{Coordination-number evolution for three imposed shear-stress levels, $\tau_1=0.90\tau_{ss}$, $\tau_2=0.68\tau_{ss}$, and $\tau_3=0.46\tau_{ss}$, at the highest pressurization rate, $R_3$, under drained and undrained hydraulic boundary conditions. The coordination number is used as a bulk measure of contact-network connectivity. The curves are primarily separated by imposed shear-stress level: the case closest to failure, $\tau_1$, has the lowest coordination number, whereas the lower shear-stress cases retain higher contact connectivity. For a fixed shear-stress level, drained and undrained cases show similar coordination-number evolution.}
  \label{fig:Coordination_number}
\end{figure}

\begin{figure*}[t]
  \centering
  \includegraphics[width=1.0\textwidth]{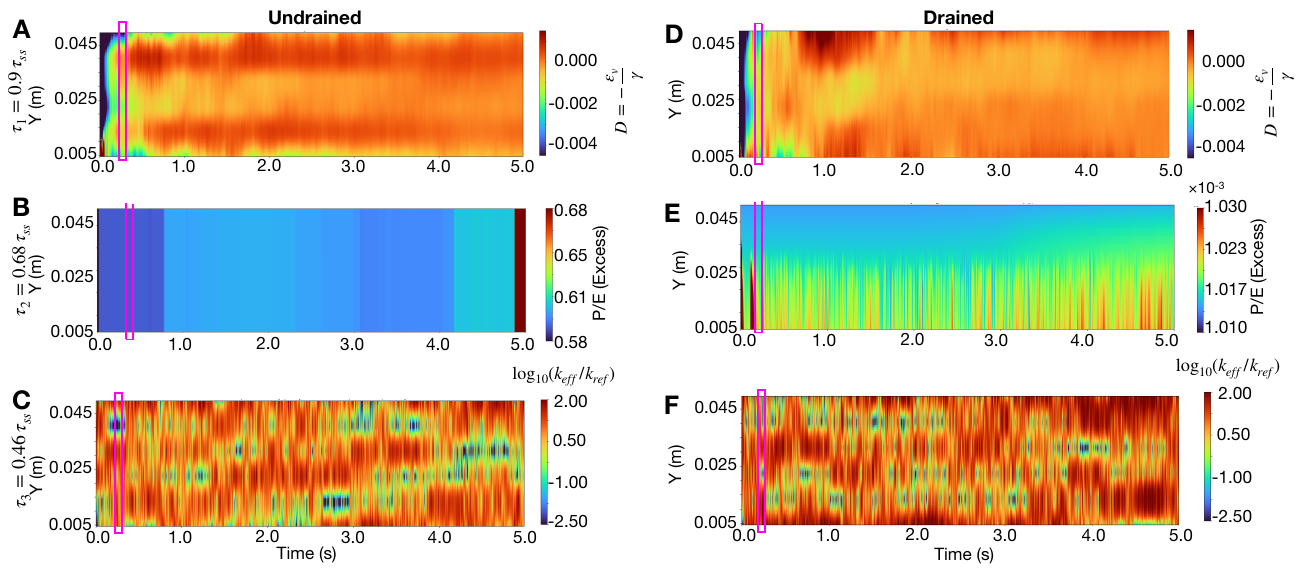}
  \caption{Representative spatiotemporal evolution of internal hydromechanical fields for pressurization-driven failure under the same loading level, $\tau_1=0.9\tau_{ss}$, comparing the undrained case (A--C, left column) and the drained case (D--F, right column). From top to bottom, the rows show cumulative dilatancy, $D=-\varepsilon_v/\gamma$, normalized excess pore pressure, $P/E$, Where,  $E=100~MPa$ is particle Young's modulus and the porosity-derived effective permeability, $\log_{10}(k_{\mathrm{eff}}/k_{\mathrm{ref}})$, plotted as functions of height and time. The purple box marks the identified failure interval. It indicates the failure interval identified from the apparent-friction time series using the threshold $\Delta\mu\ge0.01$, following the friction-drop-based slip-event criterion of Dorostkar et al.~\citep{Dorostkar2017}; the same interval is used here to relate weakening to dilatancy, pore-pressure redistribution, and permeability reorganization. In the undrained case, failure is preceded by rapid pore-pressure buildup and a comparatively uniform high-pressure state across the layer, accompanied by localized dilation/compaction structure and reorganization of the permeability field. In the drained case, pore pressure remains much smaller in magnitude and retains a persistent vertical gradient from the pressurized base toward the drained top boundary, while dilatancy and permeability evolve more gradually. The comparison shows that drainage suppresses system-wide pressurization and buffers the internal reorganization associated with failure, leading to a more spatially structured and mechanically moderated response.}
  \label{fig:Dilatancy_Porepressure_permeability}
\end{figure*}

\section{Numerical framework and pressurization-driven response}

The numerical implementation builds on the two-way coupled, volume-averaged CFD--DEM framework used in our previous work \citep{Chhushyabaga2025}, but it is applied here to a different boundary-value problem. Instead of simulating gravitational collapse and runout of initially unsupported granular masses, the present simulations impose shear stress on a confined, fluid-saturated granular layer and progressively increase basal pore pressure to trigger reactivation. This configuration is designed to isolate the coupled evolution of effective stress, slip acceleration, drainage, volumetric deformation, and contact-network degradation during pressurization-driven instability. The numerical design follows the same logic as pore-pressure-step creep and fluid-reactivation studies of granular shear zones: a reference frictional strength is first established under controlled shearing, and pore pressure is then increased under imposed shear stress so that slip rate, dilation, pressure redistribution, and contact-network degradation emerge from the coupled dynamics rather than from prescribed displacement \citep{Scuderi2017,Cappa2019,Nguyen2021,Sarma2025}. This section therefore presents the numerical framework together with the corresponding hydromechanical and micromechanical responses, so that each modeling element is introduced where it is needed to interpret the observed behavior.

\subsection{CFD--DEM formulation, model geometry, and strength calibration}

The simulations adopt a hybrid Eulerian--Lagrangian formulation. The pore fluid is solved on a fixed Eulerian grid using locally averaged fluid equations, while individual grains are tracked as Lagrangian particles whose translational and rotational motions satisfy Newton's laws. This volume-averaged CFD--DEM approach resolves particle-scale kinematics, contact forces, and fabric evolution while representing pore pressure and fluid velocity at the computational-cell scale \citep{Syamlal1993,MFIXDEM2012,Musser2021}. It is therefore distinct from pore-scale direct numerical simulation, but is well suited for dense saturated granular layers in which the macroscopic response depends on the feedback between grain rearrangement and cell-scale fluid pressure.

For an incompressible Newtonian pore fluid, the locally averaged fluid mass balance is written as
\begin{equation}
\frac{\partial \epsilon_f}{\partial t}
+\nabla\cdot(\epsilon_f \mathbf{u}_f)=0,
\end{equation}
where $\epsilon_f$ is the local fluid volume fraction and $\mathbf{u}_f$ is the averaged fluid velocity. The corresponding momentum balance can be expressed as
\begin{equation}
\frac{\partial(\epsilon_f\rho_f\mathbf{u}_f)}{\partial t}
+\nabla\cdot(\epsilon_f\rho_f\mathbf{u}_f\mathbf{u}_f)
=
-\epsilon_f\nabla p
+\nabla\cdot(\epsilon_f\boldsymbol{\tau}_f)
+\epsilon_f\rho_f\mathbf{g}
+\mathbf{M}_{sf},
\end{equation}
where $\rho_f$ is fluid density, $p$ is pore pressure, $\boldsymbol{\tau}_f$ is the viscous stress tensor, $\mathbf{g}$ is gravitational acceleration, and $\mathbf{M}_{sf}$ is the interphase momentum exchange from the solid phase to the fluid phase \citep{Syamlal1993,MFIXDEM2012,Musser2021}. The grains obey
\begin{equation}
m_i\frac{d\mathbf{v}_i}{dt}
=
\sum_j\mathbf{F}_{ij}^{c}
+\mathbf{F}_i^{f}
+m_i\mathbf{g},
\end{equation}
\begin{equation}
\mathbf{I}_i\frac{d\boldsymbol{\omega}_i}{dt}
=
\sum_j\mathbf{r}_{ij}\times\mathbf{F}_{ij}^{t},
\end{equation}
where $m_i$, $\mathbf{v}_i$, $\mathbf{I}_i$, and $\boldsymbol{\omega}_i$ are the particle mass, velocity, moment of inertia, and angular velocity. The contact force $\mathbf{F}_{ij}^{c}$ includes normal and tangential components, and $\mathbf{F}_{i}^{f}$ includes pressure-gradient and drag contributions from the pore fluid. The drag component is computed using the Syamlal--O'Brien hindered-drag closure implemented in MFIX--DEM \citep{Syamlal1993,MFIXDEM2012,Musser2021}. As in our previous CFD--DEM work\citep{Chhushyabaga2025}, only hindered drag is included; lift, virtual-mass, and explicit lubrication forces are not considered.  The pressure-gradient force acts directly on the particles, while the equal and opposite interphase force is projected back to the Eulerian grid, preserving two-way momentum coupling. This feedback is central to the present problem because pore-pressure changes modify effective confinement, while particle rearrangement and volumetric deformation modify the pore space through which pressure redistributes.

Particle--particle and particle--wall interactions are modeled using a soft-sphere Hertzian contact law with Coulomb friction \citep{CundallStrack1979,MFIXDEM2012}. This contact model permits sliding, rolling, contact loss, and force-chain reorganization during localization and failure. These micromechanical processes are essential because previous DEM studies show that accelerated slip in granular shear zones is controlled by the evolution and collapse of force-bearing contact networks rather than by bulk stress changes alone \citep{MorganBoettcher1999,Radjai1998,Tordesillas2007,Nguyen2021}. The simulations used an initial fluid time step of $\Delta t_f=10^{-3}$~s, with adaptive limits $\Delta t_{\min}=10^{-7}$~s and $\Delta t_{\max}=10^{-2}$~s. Field data were written every $10^{-3}$~s for post-processing. Particle contacts were modeled using the Hertzian DEM contact law, and MFIX--DEM subcycled the particle motion internally to resolve contact dynamics during each fluid update. This follows the time-integration approach used in our previous CFD--DEM simulations~\citep{Chhushyabaga2025} and the MFIX--DEM implementation guidance~\citep{MFIXDEM2012,MFIX_DEM_Ref,MFIX_Time}.

The numerical specimen is a dense, fluid-saturated granular layer confined between a fixed lower plate and a movable upper plate in simple shear (Fig.~\ref{fig:Model_Setup}a). Particles are generated, allowed to settle into a mechanically stable packing, and then trimmed to define the final sheared layer. The model dimensions are $L=50d_p$, $W=50d_p$, and $H\simeq38d_p$ after trimming, where $d_p\approx1$~mm is the mean particle diameter. Particle sizes follow a Gaussian distribution with standard deviation $0.1d_p$, which reduces crystallization and promotes disordered granular fabric. Periodic boundaries are applied in the horizontal directions to represent a laterally extensive shear layer and to avoid artificial sidewall constraints, following common DEM and coupled DEM treatments of granular shear zones \citep{MorganBoettcher1999,Nguyen2021}. The grain and fluid properties are summarized in Table~\ref{tab:properties}.

\begin{table}
\caption{Properties of the fluid and particles used in the CFD--DEM simulations.}
\label{tab:properties}
\centering
\begin{tabular}{l c}
\hline
Property & Value \\
\hline \rule{0pt}{2.5ex}
Fluid density, $\rho_f$ & 1000.0~kg~m$^{-3}$ \\
Fluid viscosity, $\eta$ & 0.001~Pa~s \\
Particle density, $\rho_p$ & 2500.0~kg~m$^{-3}$ \\
Particle Young's modulus, $E$ & 100~MPa \\
Particle Poisson's ratio, $\nu$ & 0.23 \\
Interparticle friction coefficient, $\mu_p$ & 0.5 \\
\hline
\end{tabular}
\end{table}

The Eulerian fluid grid was selected following the unresolved CFD--DEM grid-resolution criteria used in our previous work~\citep{Chhushyabaga2025}. For the present shear-layer domain, the minimum cell spacing was kept within the unresolved CFD--DEM grid-to-particle guideline of approximately $2$--$4d_p$, consistent with verification studies and with the grid-resolution criterion used in previous simulations~\citep{WangLiu2021,Andrews1996,Chhushyabaga2025}. Therefore, the pore-pressure, porosity, fluid-volume-fraction, and fluid--solid momentum-exchange fields reported in this study should be interpreted as Eulerian cell-averaged quantities at this unresolved CFD--DEM resolution, rather than pore-scale fields. As in our previous CFD--DEM simulations, modest mesh coarsening and refinement around the selected grid-to-particle resolution did not alter the qualitative pore-pressure evolution, failure mode, or frictional trends, although finer grids sharpen local pressure and porosity gradients~\citep{MFIXDEM2012,MFIX_DEM_Ref,MFIX_Time}.

The loading protocol contains two linked stages. In the first stage, strain-controlled simulations are performed under servo-controlled normal stress to establish the reference strength of the granular layer. Shear stresses measured at different normal stresses are fitted with a Mohr--Coulomb relation,
\begin{equation}
\tau_f=c+\mu_{\rm MC}\sigma_n ,
\end{equation}
where $c$ is the apparent cohesion and $\mu_{\rm MC}$ is the Mohr--Coulomb friction coefficient. The fitted envelope gives $c=1905.76$~Pa and a reference critical shear stress of approximately 2423~Pa for the loading condition shown in Fig.~\ref{fig:Model_Setup}b. This strain-controlled stage is used only to define the strength scale; the instability itself is examined under stress control, because imposed displacement rates can suppress the natural acceleration that characterizes failure \citep{RathbunMarone2010,Nguyen2021,Sarma2025}.

In the second stage, the control mode is changed to stress-controlled pressurization. The upper plate is servo-controlled to maintain $\sigma_n=20$~kPa and a prescribed shear stress, while pore pressure is increased from the base. Three imposed shear stresses are used,
\begin{equation}
\tau_1=0.90\tau_{ss},\qquad
\tau_2=0.68\tau_{ss},\qquad
\tau_3=0.46\tau_{ss},
\end{equation}
where $\tau_{ss}$ is the reference steady-state shear strength. Thus, $\tau_1$ represents the closest-to-failure state and $\tau_3$ the farthest-from-failure state. The imposed base pressurization rates are
\begin{equation}
R=\frac{\dot{P}_f}{\sigma_n},
\end{equation}
with $\dot{P}_f=462$, 798, and 3360~Pa~s$^{-1}$, corresponding to $R_1=0.0231$~s$^{-1}$, $R_2=0.0399$~s$^{-1}$, and $R_3=0.168$~s$^{-1}$, respectively (Fig.~\ref{fig:Model_Setup}d). The hydraulic boundary condition is varied independently: the undrained simulations retain excess pore pressure by imposing an impermeable upper boundary, whereas the drained simulations maintain atmospheric pressure at the upper boundary, allowing vertical pressure dissipation. The simulation parameters is summarized in Table~\ref{tab:simulation_cases}.

\begin{table}
\caption{Summary of stress-controlled fluid pressurization simulations. All cases were performed at a constant normal stress of $\sigma_n=20$~kPa under both drained and undrained boundary conditions.}
\label{tab:simulation_cases}
\centering
\begin{tabular}{l c}
\hline
Parameter & Values \\
\hline \rule{0pt}{2.5ex}
Loading condition & Stress-controlled \\
Imposed shear stress, $\tau/\tau_{ss}$ & 0.90, 0.68, 0.46 \\
Base pressurization rate, $\dot{P}_f$ & 462, 798, 3360~Pa~s$^{-1}$ \\
Normalized rate, $R=\dot{P}_f/\sigma_n$ & 0.0231, 0.0399, 0.1680~s$^{-1}$ \\
Hydraulic boundary condition & Drained, Undrained \\

\hline
\end{tabular}
\end{table}

\section{Drainage-controlled instability nucleation}

The failure interval highlighted in Figs.~\ref{fig:Fric_pres_un}, \ref{fig:Fric_pres_d} \ref{fig:Dilatancy_Porepressure_permeability}, and \ref{fig:Porosity} were identified objectively from the apparent-friction time series, $\mu_{\rm app}(t)$. We define failure onset as the local maximum in $\mu_{\rm app}$ immediately preceding a sustained weakening event, and the end of failure as the subsequent local minimum after the main friction drop. The friction drop is defined as
\begin{equation}
\Delta\mu =
\mu_{\rm app}^{\rm peak}-\mu_{\rm app}^{\rm min}.
\end{equation}
A candidate weakening interval was classified as failure when $\Delta\mu\ge0.01$. This threshold follows the friction-drop component of the slip-event identification procedure used by Dorostkar et al.~\citep{Dorostkar2017}, who used a friction-coefficient drop threshold of 0.01, to distinguish slip events from smaller microslip fluctuations in saturated granular fault-gouge simulations. In the present simulations, the same friction-drop threshold is applied consistently to all drainage conditions, pressurization rates, and imposed shear-stress levels. Therefore, the pre-failure, during-failure, and post-failure states are selected from the same objective weakening criterion rather than by visual inspection for analysis in Figs.~\ref{fig:Micro_Mechanics} and~\ref{fig:Force_Fabric_Anisotropy}

Figures~\ref{fig:Fric_pres_un} and \ref{fig:Fric_pres_d} show the pressurization-driven response at the highest pressurization rate, $R_3$, for the three imposed shear-stress levels. Under undrained conditions, failure occurs abruptly once retained pore pressure reduces the effective normal stress enough that the granular skeleton can no longer support the imposed shear stress (Fig.~\ref{fig:Fric_pres_un}). The weakening interval is marked by a rapid friction drop, strong excess pore-pressure buildup, and subsequent pressure redistribution or release. This behavior is consistent with effective-stress-controlled instability, but the simulations show that failure is not a homogeneous Mohr--Coulomb threshold process. Instead, pore-pressure retention, local deformation, and contact-network degradation evolve together. Similar distinctions between macroscopic failure criteria and grain-scale failure mechanisms have been emphasized in pressurized granular shear-zone simulations and pore-pressure-step creep experiments \citep{Scuderi2017,Cappa2019,Nguyen2021}.

The imposed shear-stress level controls the pressure increment required to trigger instability. At $\tau_1=0.90\tau_{ss}$, the layer is close to the reference strength and fails after a relatively small increase in pore pressure. At $\tau_3=0.46\tau_{ss}$, the layer is farther from failure and must undergo a larger hydromechanical excursion before weakening occurs. Consequently, the lower-shear case develops larger transient pore-pressure buildup and a larger frictional excursion before instability. The relevant control is therefore not absolute pore pressure alone, but the evolving distance between the current stress state and the strength of the granular skeleton. This interpretation is consistent with theoretical and experimental arguments that fluid-induced failure depends on pressure history, loading path, and stress proximity to failure \citep{Rice1975,Rudnicki1981,Wang2020,Sarma2025}.

The pore-pressure maps in Fig.~\ref{fig:Fric_pres_un} show that the approach to failure remains spatially heterogeneous even under undrained conditions. Localized pressure concentrations develop before and during the main weakening interval, indicating that layer-averaged pore pressure is insufficient to characterize where failure nucleates. These pressure heterogeneities are coupled to local volumetric deformation and evolving hydraulic connectivity. Once failure begins, the pressure field reorganizes rapidly, consistent with grain rearrangement altering the internal pore-space structure and local flow pathways \citep{YangJuanes2018,Nguyen2021}.

The drained simulations follow a different pathway (Fig.~\ref{fig:Fric_pres_d}). Weakening still occurs, but it is smoother and is accompanied by smaller excess pore-pressure amplitudes. The pressure field maintains a persistent vertical gradient from the pressurized base to the drained top boundary, showing that fluid escape continuously removes part of the imposed pressure perturbation. Drainage therefore suppresses the positive feedback between pore-pressure retention, effective-stress reduction, and contact-network unloading. Instead of rapid near-layer-wide effective-stress collapse, the drained layer weakens progressively through competition among pressurization, pressure dissipation, volumetric deformation, and granular-fabric degradation.

This drained--undrained contrast demonstrates that hydraulic boundary condition is a first-order control on instability style. Undrained conditions favor rapid pressure retention, abrupt weakening, and dynamic slip acceleration. Drained conditions preserve pressure gradients, reduce excess-pressure amplitude, and moderate the transition to failure. This distinction is directly relevant to saturated fault gouge, rainfall-triggered slopes, and injection-perturbed granular layers, where the drainage length scale and hydraulic diffusivity govern whether pore-pressure forcing produces local pressure gradients or broader effective-stress collapse \citep{SegallRice1995,Proctor2020,Brantut2020,Dunham2024}.

\section{Micromechanical fabric evolution and force-network degradation}

The macroscopic weakening described above is rooted in the evolution of the granular contact network. We therefore analyze particle rotation, force-chain structure, coordination number, and porosity profiles before, during, and after failure. These diagnostics connect the system-scale friction response to the grain-scale mechanisms that support or destroy shear resistance.

Particle rotation is used as a kinematic indicator of rolling, rearrangement, and localized shear accommodation. In Fig.~\ref{fig:Micro_Mechanics}, rotations are relatively limited before failure but become concentrated in preferred zones during the weakening interval. This transition indicates that deformation is increasingly accommodated by localized particle rolling and rearrangement rather than by homogeneous shear. Similar behavior has been observed in DEM simulations of pressurized shear zones, where slow creep is associated with distributed deformation while accelerated dynamic slip is accommodated by intense particle rotation within localized shear bands \citep{MorganBoettcher1999,Nguyen2021}.

The force-chain maps show that localized rotation coincides with degradation of the strong-force backbone. Before failure, shear resistance is carried by a connected and anisotropic network of strong contacts. As pore pressure increases and effective confinement decreases, normal contact forces are reduced, frictional resistance at contacts drops, and the force network becomes fragmented. During failure, strong chains collapse, reorient, or redistribute. This behavior is consistent with the force-chain buckling and collapse mechanisms identified in granular shear studies, where loss of a coherent strong-force network provides the micromechanical pathway to macroscopic weakening \citep{Radjai1998,Tordesillas2007,Nguyen2021}.

The representative micromechanical cases in Fig.~\ref{fig:Micro_Mechanics} further show that failure intensity depends on initial stress proximity to failure. The high-shear case, $\tau_1=0.90\tau_{ss}$, reaches instability after a smaller pressure perturbation and therefore requires less extensive fabric degradation. The lower-shear case, $\tau_3=0.46\tau_{ss}$, requires greater pressure buildup and shows stronger contact loss, force-chain reorganization, and porosity redistribution. Thus, pore pressure triggers failure by reducing the strength margin, but rapid weakening occurs only when the load-bearing contact network has degraded sufficiently.

Porosity profiles confirm that failure is not a simple uniform expansion of the layer. Instead, porosity redistributes across the layer thickness: some regions dilate and lose contacts, while others compact or maintain greater connectivity. This mixed volumetric response resembles the coexistence of dilation and compaction observed in granular shear-zone simulations and experiments, where localized deformation produces spatially variable contact density and pore-space structure \citep{RathbunMarone2010,Nguyen2021,Sarma2025}. The porosity redistribution also modifies $k_{\mathrm{eff}}$, linking mechanical fabric evolution to hydraulic reorganization.

Coordination-number evolution provides a compact measure of this contact-network degradation. The local coordination number is the number of contacts associated with each particle, and the mechanical coordination number can be written as
\begin{equation}
Z_m=\frac{2N_c}{N_p-N_r},
\end{equation}
where $N_c$ is the number of mechanically active contacts, $N_p$ is the number of particles, and $N_r$ is the number of particles that do not contribute significantly to the load-bearing network.  Figure~\ref{fig:Coordination_number} shows that the coordination number is primarily organized by the imposed shear-stress level. The case closest to failure, $\tau_1=0.90\tau_{ss}$, has the lowest coordination number, whereas the lower imposed shear-stress cases, especially $\tau_3=0.46\tau_{ss}$, retain higher contact connectivity. This indicates that proximity to failure is associated with a less connected or more degraded bulk contact network. The drained and undrained curves are similar for a fixed shear-stress level and pressurization rate. Therefore, the drainage-controlled difference in instability style is not explained by a large change in the bulk number of contacts.

To quantify the directional organization of the contact and force-bearing networks, we compute second-order anisotropy coefficients from the projected contact orientations, following the stress--force--fabric framework of Rothenburg and Bathurst~\citep{Rothenburg1989}. The contact-fabric anisotropy is defined as
\begin{equation}
a_c =
2\sqrt{
\left<\cos 2\theta\right>^2+
\left<\sin 2\theta\right>^2
},
\end{equation}
where $\theta$ is the projected contact orientation and the brackets denote averaging over contacts. This coefficient describes the directional organization of the full contact network. To account for load-bearing structure, we also compute a force-weighted anisotropy,
\begin{equation}
a_F =
2\sqrt{
\left(
\frac{\sum_k F_k\cos 2\theta_k}{\sum_k F_k}
\right)^2+
\left(
\frac{\sum_k F_k\sin 2\theta_k}{\sum_k F_k}
\right)^2
},
\end{equation}
where $F_k$ is the force-chain intensity associated with contact $k$. The strong-contact anisotropy $a_s$ is computed using only contacts whose force-chain intensity exceeds the selected strong-force threshold. Thus, $a_c$ describes the full contact fabric, $a_F$ describes the force-weighted load-bearing fabric, and $a_s$ describes the directional organization of the strong-force backbone.

Figure~\ref{fig:Force_Fabric_Anisotropy} compares these anisotropy measures for the representative undrained R3--T1 and R3--T3 cases. The left three columns show R3--T1, which is closest to the reference shear strength, whereas the right three columns show R3--T3, the lowest-shear case at the same pressurization rate. Within each case, the columns correspond to pre-failure, during-failure, and post-failure states. From top to bottom, the rows show $a_s$ (panels A and D), $a_F$ (panels B and E), and $a_c$ (panels C and F).

The two cases exhibit distinct post-failure fabric evolution. In R3--T1, the anisotropy measures remain relatively persistent through failure: $a_s$, $a_F$, and $a_c$ remain close to their pre-failure values after failure. This indicates that the high-shear layer, which begins close to the reference strength, can undergo pore-pressure-triggered weakening while retaining a coherent anisotropic load-bearing fabric. In contrast, R3--T3 shows a pronounced post-failure reduction in all three measures, with $a_c$ and $a_F$ decreasing to nearly isotropic values and $a_s$ also decreasing substantially. This indicates stronger degradation of both the geometric contact fabric and the strong-force backbone in the lower-shear case.

These trends complement the spatial micromechanical fields shown in Fig.~\ref{fig:Micro_Mechanics}. While Fig.~\ref{fig:Micro_Mechanics} identifies where particle rotation, force-chain reorganization, coordination changes, and porosity redistribution occur, Fig.~\ref{fig:Force_Fabric_Anisotropy} quantifies how the directional structure of the contact and force-bearing networks evolves during the same failure sequence. The consistently larger values of $a_s$ relative to $a_F$ and $a_c$ show that the strong-force backbone is more directionally organized than the full contact network. The sharp post-failure reduction of all three measures in R3--T3 indicates that macroscopic weakening is associated with loss of directional load-bearing structure, not only with changes in contact number.

The anisotropy comparison also links the micromechanical response to the macroscopic behavior discussed in Figs.~\ref{fig:Fric_pres_un} and~\ref{fig:Fric_pres_d}. R3--T1 starts closer to failure and weakens after a smaller pressure perturbation while maintaining a relatively coherent force-bearing fabric. R3--T3 starts farther from failure, requires larger pore-pressure buildup, and undergoes stronger post-failure fabric degradation. Thus, stress proximity to failure controls not only the pressure increment required for instability, but also the degree of force-network degradation needed before the granular skeleton can no longer support the imposed shear stress.

These observations support a mechanistic sequence for pressurization-driven failure. Pore-pressure increase reduces effective confinement; reduced confinement weakens contact forces; localized particle rotation and rearrangement intensify; force chains collapse or reorient; coordination number decreases; and the granular skeleton loses the connectivity required to sustain the imposed shear stress. Macroscopic weakening is therefore the system-scale expression of contact-network degradation.

\section{Dilatancy, pore-pressure structure, and hydraulic reorganization}

To connect macroscopic weakening to internal hydromechanical structure, we coarse-grain particle data onto the Eulerian grid to obtain local solid fraction, porosity, velocity, strain rate, and pore pressure. The local solid fraction is
\begin{equation}
\phi_s(\mathbf{x},t)=\frac{1}{V_c}\sum_{i\in V_c}V_i,
\end{equation}
where $V_i$ is the volume of particle $i$ contributing to cell volume $V_c$ and corresponding Eulerian averaging volume is the CFD cell volume, $V_c$, used in the locally averaged fluid equations and in the fluid--solid momentum-exchange term. Following our previous work~\citep{Chhushyabaga2025}, $V_c$ is defined from the Eulerian cell dimensions as $V_c=\Delta x_f\Delta y_f\Delta z_f$, where $\Delta x_f$, $\Delta y_f$, and $\Delta z_f$ are the fluid-cell dimensions in the streamwise, transverse, and vertical directions, respectively. Equivalently, the averaging volume may be expressed as $V_c=(\Delta x_f/d_p)(\Delta y_f/d_p)(\Delta z_f/d_p)d_p^3$. With $d_p\simeq1$~mm, this corresponds to $V_c\simeq(\Delta x_f/d_p)(\Delta y_f/d_p)(\Delta z_f/d_p)\times10^{-9}$~m$^3$. Thus, the pore-pressure, porosity, fluid-volume-fraction, and fluid--solid momentum-exchange fields reported in this study represent Eulerian cell-averaged quantities over the CFD cell volume, rather than pore-scale values. Porosity is then
\begin{equation}
n(\mathbf{x},t)=1-\phi_s(\mathbf{x},t).
\end{equation}
Local velocity gradients are reconstructed using a least-squares procedure \citep{Rycroft2009,Chhushyabaga2025}, and cumulative dilatancy is computed as
\begin{equation}
D=-\frac{\varepsilon_v}{\gamma},
\end{equation}
where positive $D$ denotes dilation and negative $D$ denotes compaction. This convention allows direct comparison between pore-pressure evolution and volumetric deformation, which is essential because dilation tends to reduce pore pressure and stabilize slip, whereas compaction tends to increase pore pressure and promote weakening \citep{Frank1965,Scholz1973,SegallRice1995,RudnickiChen1988,Sarma2025}.

Figure~\ref{fig:Dilatancy_Porepressure_permeability} compares cumulative dilatancy, normalized excess pore pressure, and a porosity-derived effective-permeability proxy for representative drained and undrained simulations at $\tau_1=0.90\tau_{ss}$. The permeability proxy is defined as
\begin{equation}
k_{\mathrm{eff}}(n)=
\frac{d_p^{2}}{C_K}
\frac{n^{3}}{(1-n)^{2}},
\label{eq:keffKC}
\end{equation}
and reported as
\begin{equation}
\log_{10}\left(\frac{k_{\mathrm{eff}}}{k_{\mathrm{ref}}}\right),
\end{equation}
where $k_{\mathrm{ref}}$ is the initial depth-median value. This quantity is not a direct CFD flux--gradient permeability measurement; it is an interpretive proxy for relative hydraulic-connectivity changes associated with porosity reorganization. This distinction matters because porosity--permeability closures are useful in reduced hydromechanical models but become uncertain near failure, when fabric anisotropy, localization, and pore-space connectivity evolve heterogeneously \citep{Segall2010,Cappa2019,Sarma2025}.

In the undrained case, the dilatancy field develops alternating zones of dilation and compaction before and during the failure interval. This shows that deformation is internally partitioned rather than homogeneous. The pattern is consistent with granular shear-zone studies in which pressurization and shear drive a transition from distributed deformation toward localized or banded structures \citep{Sulem1995,RathbunMarone2010,Nguyen2021}. Despite this volumetric heterogeneity, the pore-pressure field becomes high across much of the layer near failure because the impermeable upper boundary prevents efficient pressure dissipation. Local dilation and compaction modulate the pressure field, but pressure retention dominates the layer-scale effective-stress reduction.

The porosity-derived permeability proxy reorganizes rapidly during the undrained failure interval. Zones of enhanced and reduced $k_{\mathrm{eff}}$ emerge in concert with the dilatancy structure, indicating that hydraulic connectivity evolves with granular deformation. This behavior supports a coupled feedback: pressurization weakens the granular skeleton, deformation localizes, local volumetric strain reorganizes pore space, and the evolving hydraulic structure controls how pressure anomalies are retained or redistributed. Such strongly coupled near-failure heterogeneity is difficult to represent using fixed or smoothly varying continuum permeability closures.

In the drained case, pore-pressure amplitudes remain smaller and a persistent vertical gradient develops from the pressurized base toward the drained top boundary. Dilatancy and $k_{\mathrm{eff}}$ still evolve, indicating that localization and volumetric reorganization remain intrinsic to the granular response, but their hydraulic consequences are buffered by fluid escape. Drainage therefore does not eliminate localization; it changes how localization couples to pore-pressure evolution. The same granular layer fails through broad pressure retention in the undrained case and through a more spatially structured, gradient-controlled pathway in the drained case.

\begin{figure*}[t]
  \centering
  \includegraphics[width=0.9\textwidth]{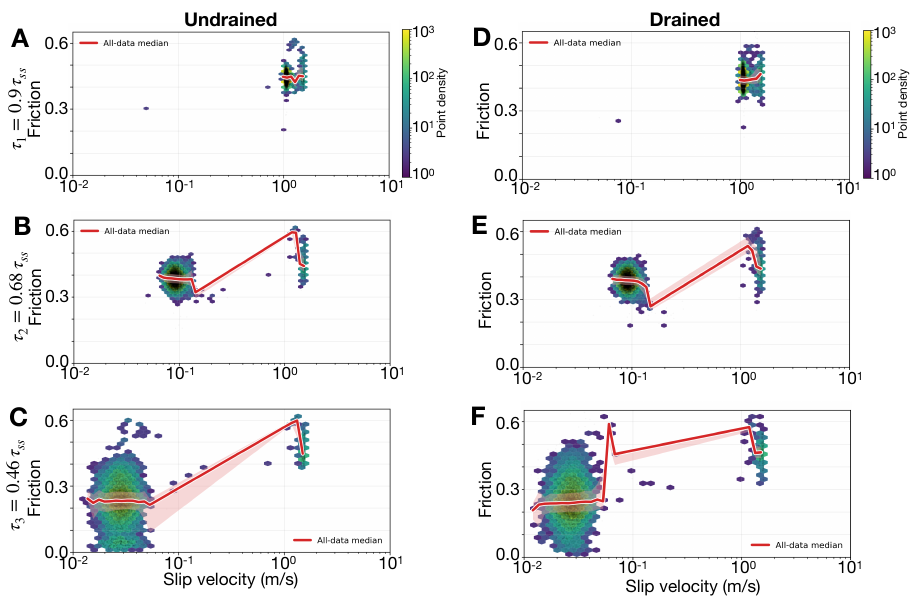}
  \caption{Friction--slip-velocity for the three pressurization rates, shown separately for the undrained case (A--C, left column) and the drained case (D--F, right column) at the same imposed shear-stress level, with rows corresponding to $\tau_1=0.9\tau_{ss}$, $\tau_2=0.68\tau_{ss}$, and $\tau_3=0.46\tau_{ss}$. In each panel, the colored background represents the point density of $(V,\mu)$ data in log-scaled slip-velocity.  This density-based visualization emphasizes the dominant frictional states and the spread of the response. Across both drainage conditions, the data cluster at low slip velocities in a quasi-steady strengthening regime and then transition toward higher velocities associated with weakening and failure. The $\tau_2$ and $\tau_3$ cases exhibit broader low-velocity clusters and a more pronounced non-monotonic median trend, whereas the $\tau_1$ case remains concentrated near the high-velocity branch. }
  \label{fig:Slip_velocity2}
\end{figure*}

\begin{figure*}[t]
  \centering
  \includegraphics[width=0.9\textwidth]{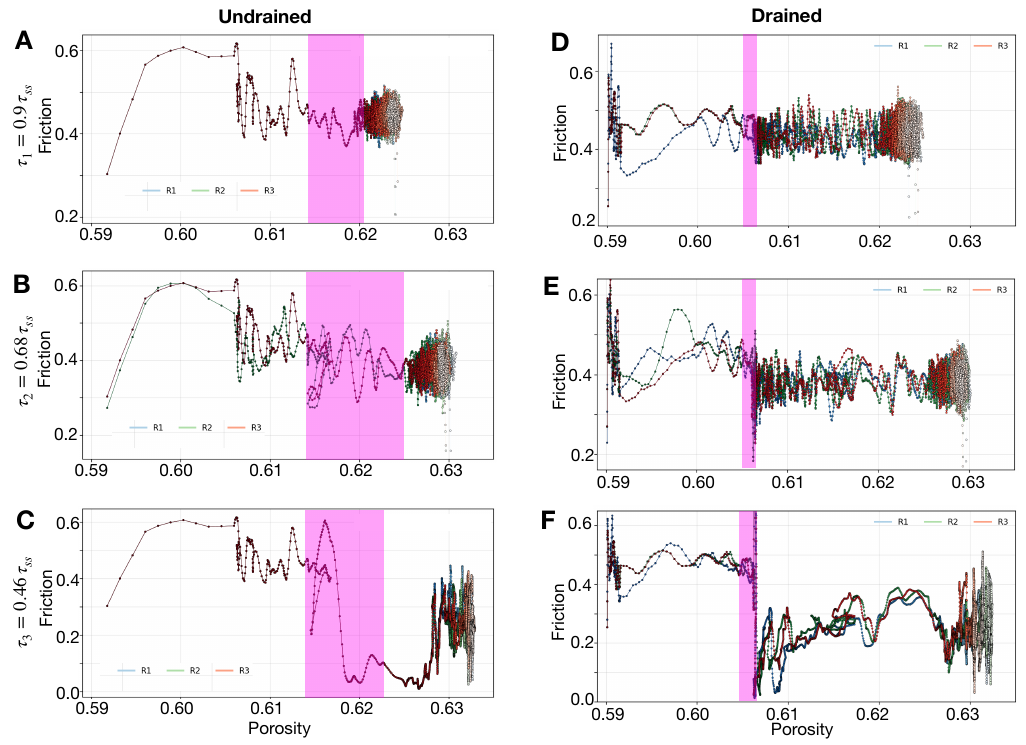}
  \caption{Friction--porosity evolution for the three pressurization rates, shown separately for the undrained case (A--C, left column) and the drained case (D--F, right column) at the same imposed shear-stress level, with rows corresponding to $\tau_1=0.9\tau_{ss}$, $\tau_2=0.68\tau_{ss}$, and $\tau_3=0.46\tau_{ss}$. In each panel, friction is plotted against bulk porosity, and the purple band marks the porosity range associated with the identified failure window. It marks the range of porosity values sampled during the objectively identified failure interval, defined from the peak-to-minimum friction drop in $\mu_{\rm app}(t)$ using the threshold $\Delta\mu\ge0.01$~\citep{Dorostkar2017}. The curves are colored by pressurization rate ($R_1$, $R_2$, and $R_3$), allowing direct comparison of rate effects at fixed drainage condition and shear-stress level. In both drainage conditions, friction initially increases with porosity during dilatancy-driven strengthening, indicating that shear-induced expansion of the granular layer is accompanied by frictional hardening. As the system approaches instability, the trajectories enter the highlighted failure range and friction drops sharply with only limited additional porosity increase, consistent with rapid weakening triggered by collapse or reorganization of the load-bearing contact network within an already dilated layer. The undrained case exhibits larger and more abrupt excursions through the failure range, whereas the drained case shows smoother, more distributed trajectories, indicating that drainage buffers the frictional response and moderates the porosity pathway to failure.}
  \label{fig:Porosity}
\end{figure*}

\section{Slip dynamics, rate dependence, and porosity-dependent weakening}

Figures~\ref{fig:Slip_velocity2} and \ref{fig:Porosity} show how the coupled hydromechanical and micromechanical processes appear in friction--velocity and friction--porosity space for the full set of pressurization rates and shear-stress levels. These projections should not be interpreted as calibrated single-valued constitutive laws. Rather, they show the regions of state space occupied by the layer as pore pressure, slip velocity, porosity, drainage condition, and contact fabric evolve together.

The friction--velocity response contains a low-velocity regime associated with gradual deformation and a higher-velocity regime associated with weakening and instability. At low velocities, the apparent frictional response is consistent with dilatancy-dominated strengthening: incremental shear is accompanied by volumetric expansion, particle rearrangement, and mechanical work against confinement. This is consistent with classical granular dilatancy and with saturated-gouge studies showing that dilation can temporarily stabilize slip by reducing pore pressure and increasing effective normal stress \citep{Reynolds1885,Frank1965,Marone1991,SegallRice1995,Sarma2025}. As pressurization continues, pore-pressure-driven unloading overwhelms this stabilizing tendency, and the system transitions toward higher velocity and weakening.

The point-density structure in Fig.~\ref{fig:Slip_velocity2} shows that the response is path dependent. The data form clusters rather than collapsing onto a single trajectory. The $\tau_2$ and $\tau_3$ cases occupy broader low-velocity regions because they begin farther from failure and therefore spend more time in the competition regime between dilatancy-driven strengthening and pore-pressure-driven weakening. The $\tau_1$ cases, being closer to failure, transition more directly toward the high-velocity weakening branch. This behavior is consistent with fluid-injection DEM results showing that pore pressure controls the approach to sliding while porosity and coordination number govern the detailed path through friction--velocity space \citep{Sarma2025}.

Drainage changes the transition between these regimes. In undrained simulations, pressure retention promotes a sharper movement from low-velocity deformation to rapid weakening. In drained simulations, pressure dissipation narrows the state-space excursions and smooths the median trends. Drainage therefore acts as a hydraulic damping mechanism: it does not prevent weakening, but it reduces the rate at which pore-pressure forcing is converted into effective-stress loss and slip acceleration.

The friction--porosity response in Fig.~\ref{fig:Porosity} provides a complementary interpretation. In both drainage conditions, friction initially increases with porosity, indicating that early dilation is associated with transient strengthening. Near failure, however, friction drops sharply with limited additional porosity increase. This means that failure is not caused simply by continued dilation. Instead, the layer reaches a state in which the already reorganized fabric can no longer sustain the imposed shear stress under reduced effective confinement.

Porosity therefore has a dual role. It is a hydraulic variable because it controls pore volume and the porosity-derived permeability proxy; it is also a mechanical state variable because it reflects packing structure and contact density. During early deformation, dilation can strengthen the layer through dilatant hardening. Near failure, the same dilation can move the system toward a looser and less connected fabric, making it more susceptible to force-chain collapse when pore pressure remains high or continues to increase. The failure pathway is therefore governed by competition between stabilizing dilatant hardening and destabilizing fabric degradation, rather than by a monotonic dependence on porosity alone \citep{Sarma2025}.

\section{Comparison of Frictional Rheology}

To place the pressurization-driven response in the context of dense-suspension and immersed-granular rheology, we compare the apparent friction coefficient with the viscous number $I_v$,
\begin{equation}
I_v = \frac{\eta_f \dot{\gamma}}{P'} .
\end{equation}
This comparison is used as a diagnostic measure to locate the CFD--DEM response relative to the experimental datasets of Boyer et al.~\citep{Boyer2011} and Houssais et al.~\citep{Houssais2016}, rather than as validation of a local $\mu(I_v)$ rheology.

For each simulation, $I_v(t)$ was computed at the each time step. The fluid viscosity was taken as $\eta_f=1.0\times10^{-3}$ Pa\,s, corresponding to water. The representative shear rate $\dot{\gamma}(t)$ was obtained from the coupled shear-rate which represents the instantaneous shear rate, rather than a local shear-band value or a time-windowed average.

The particle pressure $P'(t)$ was evaluated from the contact-force contribution carried by the granular skeleton, following the same contact-force-based stress interpretation used in our previous work~\citep{Chhushyabaga2025}. In that framework, the granular stress is obtained from interparticle contact forces within representative volume elements, and the resulting normal and shear stress components are used to define friction and viscous-number. Here, because the pressurization simulations are analyzed and we compute the representative particle pressure from the summed normal collision/contact force transmitted to the boundary,
\begin{equation}
P'(t)=\frac{|F_n^c(t)|}{A},
\end{equation}
where $F_n^c(t)$ is the summed normal collision force  and $A=0.05\times0.05~\mathrm{m}^2$ is the area of top plate. Thus, $P'$ represents the normal stress carried by the solid contact skeleton. 

The friction coefficient $\mu(t)$ was taken from the same instantaneous coupled shear-rate time series used to compute $\dot{\gamma}(t)$. The resulting output were then combined across all drained and undrained simulations, including three imposed shear-stress levels, $\tau_1=0.90\tau_{ss}$, $\tau_2=0.68\tau_{ss}$, and $\tau_3=0.46\tau_{ss}$, and three pressurization rates.

\begin{figure}[t]
  \centering
  \includegraphics[width=\columnwidth]{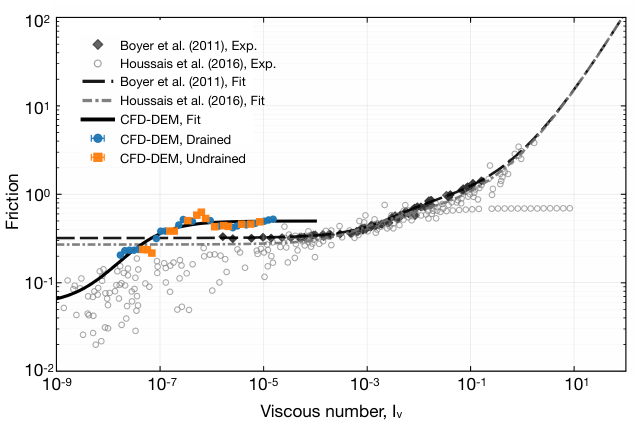}
\caption{Friction--viscous-number comparison for drained and undrained CFD--DEM pressurization simulations against dense-suspension and sediment-bed rheology \citep{Boyer2011,Houssais2016}. The apparent friction coefficient is plotted against the viscous number, $I_v=\eta_f\dot{\gamma}/P'$, where $\eta_f$ is the pore-fluid viscosity, $\dot{\gamma}$ is the representative layer-scale shear strain rate, and $P'$ is the contact-force-based particle pressure carried by the granular skeleton. Symbols denote globally binned CFD--DEM data from simulations spanning three imposed shear-stress levels, three pressurization rates, and both hydraulic boundary conditions. The CFD--DEM data occupy a very-low-viscous-number range, primarily $10^{-8}\lesssim I_v\lesssim10^{-5}$, and are fit empirically using $\mu=\mu_1+(\mu_2-\mu_1)I_v/(I_v+I_0)$, with $\mu_1=0.0565$, $\mu_2=0.498$, $I_0=4.66\times10^{-8}$, and $R^2=0.7771$. The Boyer et al.~\citep{Boyer2011} and Houssais et al.~\citep{Houssais2016} curves are shown as reference dense-suspension and sediment-bed trends, respectively. Houssais et al.~\citep{Houssais2016} showed that local Boyer-type rheology organizes their measurements mainly over $3\times10^{-5}\lesssim I_v\lesssim2$; below this range, the response enters a creeping regime in which friction does not collapse onto a unique local $\mu(I_v)$ law. The present CFD--DEM data fall mostly within this low-$I_v$ regime, so the fitted curve should be interpreted as a diagnostic summary of the binned trend rather than validation of a local viscous-number rheology.
}
  \label{fig:Friction_viscous}
\end{figure}

The binned dataset used in Fig.~\ref{fig:Friction_viscous} was constructed from instantaneous values of $I_v(t)$ and $\mu(t)$ computed for each simulation case. Because $I_v$ is strictly positive and spans several orders of magnitude, the combined dataset was binned using 35 logarithmically spaced intervals in $I_v$. This approach produces equal-width classes on the logarithmic scale, which is appropriate for variables distributed over broad positive ranges \citep{Cox2018}. A bin was retained only if it contained at least five valid instantaneous samples. For each accepted bin, the representative viscous number is the geometric bin center, while the plotted friction coefficient is the arithmetic mean of all $\mu(t)$ values within that bin. The numbers of drained and undrained samples in each bin were also tracked, allowing each bin to be classified as drained- or undrained-dominated. This procedure provides a reproducible global comparison between the transient CFD--DEM response and experimental $\mu(I_v)$ datasets.

The globally binned CFD--DEM data over $10^{-9}\le I_v\le10^{-4}$ were fit with a dense-suspension-style expression,
\begin{equation}
\mu=\mu_1+(\mu_2-\mu_1)\frac{I_v}{I_v+I_0}.
\end{equation}
The fitted parameters are $\mu_1=0.0565$, $\mu_2=0.498$, and $I_0=4.66\times10^{-8}$, with $R^2=0.7771$ for the accepted binned points. These fitted values provide an empirical description of the globally binned trend only. They should not be interpreted as validation of a local $\mu(I_v)$ law, because the data combine transient, spatially heterogeneous, accelerating states from different drainage conditions, stress levels, pressurization rates, and stages of instability.

The sediment-transport experiments of Houssais et al.~\citep{Houssais2016} provide a more direct but less homogeneous comparison for the present simulations. In that study, a bed of settling particles was sheared from above in a laminar annular flume, and refractive-index-matched imaging was used to recover depth-resolved particle concentration and velocity profiles. The particle pressure was not externally imposed, but was inferred from the submerged weight of the overlying grains, with an additional surface-pressure correction $P_0\simeq0.1(\rho_p-\rho_f)gd$ introduced to account for the finite normal load required to entrain a surface particle. Thus, Houssais et al.~\citep{Houssais2016} evaluated a local sediment-bed rheology using $\mu(z)=\tau(z)/[P_p(z)+P_0]$ and $I_v(z)=\eta_f|\dot{\gamma}(z)|/[P_p(z)+P_0]$.

For comparison, the Boyer et al.~\citep{Boyer2011} dense-suspension relation shown in Fig.~\ref{fig:Friction_viscous} uses the parameters $\mu_1=0.32$, $\mu_2=0.70$, $I_0=5.0\times10^{-3}$, and $\phi_m=0.585$ in the form
\begin{equation}
\mu=\mu_1+(\mu_2-\mu_1)\frac{I_v}{I_0+I_v}+I_v+2.5\phi_m\sqrt{I_v}.
\end{equation}
The fitted Houssais et al.~\citep{Houssais2016} curve uses the same functional form, with $\mu_1=0.27$, $\mu_2=0.52$, $I_0=1.2\times10^{-3}$, and $\phi_m=0.589$. Their measurements collapsed onto a Boyer-type local rheology only over the range $3\times10^{-5}\lesssim I_v\lesssim2$; below this range, the friction coefficient did not approach a constant quasi-static value, but instead decreased continuously with decreasing $I_v$, marking a transition to a creeping regime that cannot be described by the local model alone.

The comparison in Fig.~\ref{fig:Friction_viscous} should therefore be interpreted as a diagnostic test of rheological organization, rather than as a strict validation of the Boyer constitutive law. The partial ordering of the CFD--DEM data with increasing $I_v$ indicates that viscous-number scaling captures an important first-order control: as the ratio of viscous forcing to effective particle pressure increases, the granular skeleton becomes progressively easier to rearrange and the macroscopic friction evolves toward the dense-suspension trend. However, the low-$I_v$ response is governed by enduring multi-particle contacts, force-chain stability, intermittent rearrangements, and nonlocal transmission of stress through the contact network. In this regime, deformation occurs through grain rearrangements embedded within a mostly quasi-static contact skeleton, so a purely local relation between $\mu$ and $I_v$ is not expected to produce a unique collapse.

\section{Conclusion}

This study used two-way coupled, volume-averaged CFD--DEM simulations to investigate how fluid pressurization drives instability in a saturated granular layer. By first calibrating the reference strength under strain-controlled shear and then applying pore-pressure forcing under stress-controlled loading, the simulations isolate the mechanisms by which slip velocity, pore pressure, volumetric deformation, and contact-network evolution emerge from internal hydromechanical feedbacks rather than imposed displacement.

The results show that pressurization-driven failure is governed by the coupled evolution of effective-stress reduction, dilation or compaction, hydraulic-connectivity changes, strain localization, and degradation of the load-bearing contact skeleton. Pore pressure provides the destabilizing effective-stress forcing, but rapid failure occurs only when the granular fabric has weakened sufficiently that the imposed shear stress can no longer be supported. Layers closer to the calibrated shear strength fail after smaller pore-pressure increases, whereas layers farther from failure require larger pressure buildup and more extensive fabric reorganization before instability develops.

Drainage condition strongly controls the failure pathway. Undrained boundaries promote pressure retention, rapid excess-pressure buildup, abrupt frictional weakening, and dynamic slip acceleration. Drained boundaries maintain vertical pressure gradients, reduce excess-pressure amplitudes, and produce smoother, more progressive weakening. This contrast shows that drainage regulates how efficiently pore-pressure forcing is converted into effective-stress loss, contact-network degradation, and slip acceleration, consistent with the broader competition between pressure generation, hydraulic diffusion, and deformation-induced pore-volume change \citep{SegallRice1995,Proctor2020,Brantut2020,Sarma2025}.

The micromechanical diagnostics identify the grain-scale origin of the macroscopic instability. Alternating dilation and compaction bands reorganize porosity and the porosity-derived effective-permeability, indicating that hydraulic pathways evolve during failure. At the same time, localized particle rotation, coordination-number evolution, and collapse or redistribution of strong force chains mark progressive degradation of the anisotropic load-bearing fabric. The coordination-number analysis shows that bulk contact connectivity is controlled primarily by the imposed shear-stress level: cases closer to the reference shear strength exhibit lower coordination numbers, whereas cases farther from failure retain a more connected contact network. At fixed pressurization rate and imposed shear stress, drained and undrained simulations exhibit similar coordination-number evolution, indicating that the contrasting failure styles are not caused by a large difference in the total number of contacts. Instead, drainage controls the instability pathway mainly by regulating pore-pressure retention or dissipation, effective-stress evolution, pressure-gradient development, and force-chain reorganization. Thus, the transition to rapid slip is controlled not by bulk pore pressure alone, but by the coupled evolution of pressure, volumetric deformation, hydraulic connectivity, and granular fabric \citep{Radjai1998,Tordesillas2007,Nguyen2021}. Overall, the anisotropy results demonstrate that pressurization-driven failure is not governed only by contact loss or pore-pressure increase, but also by degradation of the directional load-bearing fabric. This degradation is strongest in the lower-shear case, whereas the near-failure high shear case weakens while retaining a more coherent anisotropic force-bearing network.

A comparison with viscous-number scaling shows that the simulated frictional response is partly organized by viscous number $I_v$, but does not collapse onto a unique local dense-suspension rheology. This behavior is consistent with the low-$I_v$ creeping regime, where intermittent rearrangement, nonlocal stress transmission, and progressive degradation of the contact network remain central to the macroscopic frictional response.

These findings provide a grain-resolving insights for interpreting pressurization-driven instability in fault gouge, rainfall-triggered slopes, debris-flow initiation, and injection-induced shear failure. Poorly drained granular layers are expected to favor abrupt weakening when local fabric degradation occurs within broader pore-pressure retention, whereas better-drained systems may fail more progressively under persistent pressure gradients. Future work should extend the framework toward direct flux-based permeability estimates, broader grain-size and shape distributions, grain breakage, mixed drainage conditions, and more realistic fault or slope geometries.

\section*{Author Declarations}
The authors declare no conflicts of interest.

\section*{Acknowledgments}
The research has been supported by start-up funds from the Department of Civil and Environmental Engineering and the Cullen College of Engineering at the University of Houston, as well as the American Chemical Society (ACS) Petroleum Research Fund under ACS-PRF Grant 66470-DNI8.

\section*{Data Availability}
The MFIX simulation scripts and open-source code used in this study are available on Zenodo under the title: \textit{MFIX Simulation Scripts and Open Source Code for research paper entitled 'Investigating frictional instability due to pressurization in granular media: insights from coupled computational fluid dynamics–discrete element method'}. The dataset can be accessed at: 
\href{https://doi.org/10.5281/zenodo.20437747}{https://doi.org/10.5281/zenodo.20437747}.

% If in two-column mode, this environment will change to single-column format so that long equations can be displayed. 
%\begin{widetext}
%$$\mbox{put long equation here}$$
%\end{widetext}

% Figures should be put into the text as floats. 
% Use the graphics or graphicx packages (distributed with LaTeX2e).
% Put the label that you will use with \ref{} command in the braces of the \label{} command.
%
% \begin{figure}
% \end{figure}

% \begin{table}
% \end{tabular}
% \end{table}

%\begin{acknowledgments}
% Put your acknowledgments here.
%\end{acknowledgments}

\section*{References}
\bibliographystyle{aipnum4-1}
\bibliography{bilbography}

\end{document}